\documentclass{article}
\usepackage{ltwol2e}

\usepackage{graphicx}
\usepackage{epsfig}

\def\Journal#1#2#3#4{{#1} {\bf #2}, #3 (#4)}

\def\NPB{{\em Nucl. Phys.} B}
\def\PLB{{\em Phys. Lett.}  B}

\def\PRD{{\em Phys. Rev.} D}
\def\ZPC{{\em Z. Phys.} C}

\def\ra{\rightarrow}

\def\be{\begin{equation}}
\def\ee{\end{equation}}
\def\bea{\begin{eqnarray}}
\def\eea{\end{eqnarray}}
\def\z{Z^0}

\def\myfigure#1#2#3#4#5#6#7{
  \begin{figure}[hbtp]
  \hskip #1
  \vskip #2
    \begin{center}
      \epsfxsize=#3
      \leavevmode
      \epsffile{#4}
    \end{center}
    \vskip #5
    \caption{#6}
    \label{#7}
  \end{figure} }

\newcommand{\as}{\alpha_s}

\newcommand{\mb}{\overline{m}_b}

\begin{document}
\begin{titlepage}
\renewcommand{\thefootnote}{\fnsymbol{footnote}}
\begin{flushright}
 FTUV/98-79 \\ IFIC/98-80
\end{flushright}
\par \vspace{10mm}
\begin{center}
{\Large \bf
NLO CALCULATIONS OF THE THREE-JET HEAVY QUARK PRODUCTION
IN $e^+e^-$-ANNIHILATION: \\
\vspace{1mm}
STATUS AND APPLICATIONS\footnote{to be published in the Proceedings of the XXIX
International Conference on High Energy Physics, 
Vancouver, B.C. Canada, July 23-29, 1998.}}
\end{center}
\par \vspace{2mm}
\begin{center}
Mikhail Bilenky$^a$\footnote{On leave from JINR, 141980 
Dubna, Russian Federation},
Germ\'an Rodrigo$^b$\footnote{On leave from Departament de
F\'{\i}sica Te\`orica, IFIC,
CSIC-Universitat de Val\`encia, 46100 Burjassot, Val\`encia, Spain} and
Arcadi Santamaria$^c$ 
\vspace*{0.5cm} \\
$^a$~Institute of Physics, AS CR, 18040 Prague and \\
Nuclear Physics Institute, AS CR, 25068 \v{R}e\v{z}(Prague), Czech Republic \\
$^b$~INFN-Sezione di Firenze, Largo E. Fermi 2, 50125 Firenze, 
Italy \\
$^c$~Departament de F\'{\i}sica Te\`orica, IFIC,
CSIC-Universitat de Val\`encia, 46100 Burjassot, Val\`encia, Spain
\end{center}
\par \vspace{2mm}
\begin{center} {\large \bf Abstract} \end{center}
\begin{quote}
Next-to-leading order 
calculations for heavy quark three-jet production in e+e- annihilation
are  reviewed. Their applications for the measurement of the b-quark
mass at LEP/SLC and for the test of flavour independence of the strong coupling 
constant are discussed.  
\end{quote}
\vspace*{\fill}
\end{titlepage}
\newpage\addtocounter{footnote}{-3}
\pagestyle{empty}

\title{NLO CALCULATIONS OF THE THREE-JET HEAVY QUARK PRODUCTION
IN $e^+e^-$-ANNIHILATION: STATUS AND APPLICATIONS}

\author{M.~BILENKY}

\address{Institute of Physics, AS CR, 18040 Prague and  
Nuclear Physics Institute, AS CR, 25068 \v{R}e\v{z}(Prague),
Czech Republic
}

\author{G.~RODRIGO}

\address{INFN-Sezione di Firenze, Largo E. Fermi 2, 50125 Firenze, 
Italy
}  

\author{A.~SANTAMARIA}

\address{Departament de F\'{\i}sica Te\`orica, IFIC,
CSIC-Universitat de Val\`encia, 46100 Burjassot, Val\`encia, Spain
}  

\twocolumn[\maketitle\abstracts{  
Next-to-leading order 
calculations for heavy quark three-jet production in e+e- annihilation
are  reviewed. Their applications for the measurement of the b-quark
mass at LEP/SLC and for the test of flavour independence of the strong coupling 
constant are discussed.  
}]

\section{Motivation}
Effects of the bottom-quark mass,
$m_b$, have been already noticed in the early tests~\cite{lep} 
of the flavour independence of 
the strong coupling constant, $\as$, in $e^+e^-$-annihilation 
at the $\z$-peak. 
They became very significant
in the final analysis, which included millions of hadronic 
$\z$-decays~\cite{delphi97,chrisman,sld98,burrows}. For example, if
the b-quark mass is neglected, the ratio $\as^b/\as^{light}$, where
$\as^b$ measured from hadronic events with b-quarks in the final state,
and $\as^{light}$ from events with light quarks ($uds$), is
shifted from one~\cite{chrisman} by $8\%$.
Thus, high LEP/SLC precision
requires an accurate account for the heavy quark mass\footnote{It is mainly
related to the bottom-quark. Effects of the charm-quark mass are smaller, roughly
by the factor $m_c^2/m_b^2$.}
in the theoretical predictions
for the $e^+e^-$-annihilation into jets at the $\z$-pole. 

The quark mass effects in the $\z$ decays were discussed in 
the literature~\cite{rev}.
The leading order (LO) complete Monte-Carlo calculation for $e^+e^- \ra 
3jets, 4jets$ with massive quarks was first done in~\cite{bmm92}. 
Later, motivated by the remarkable sensitivity of the three-jet observables
to the value of the quark mass, the possibility of the determination of $m_b$
at LEP, assuming universality of strong interactions, was considered~\cite{juan,brs95}.
This question was analyzed in detail in~\cite{brs95}, where
the necessity of the next-to-leading
order (NLO) calculation for the measurements of the $m_b$ was also emphasized.

The NLO calculations for the
process $e^+e^- \ra 3jets$, with complete effects of the quark mass, were
performed independently by three groups \cite{rsb97,bbu97,no97}.
These predictions are in agreement with each other
and were successfully used in the 
measurements of the b-quark mass far above threshold~\cite{delphi97,burrows}
and in the precision tests of the universality of the strong 
interactions~\cite{delphi97,chrisman,sld98,burrows} at the 
$\z$-pole\footnote{Due to the large correlation between $\as$ and $m_b$,
either $m_b$ or $\as$ was treated as a free parameter.
The value of another one was taken from other measurements.}.

In this talk we make a short review of next-to-leading order predictions
for $e^+e^- \ra 3 jets$ including effects of the quark mass 
and its applications at 
the $\z$-peak\footnote{We would like to note that elements of
these calculations can be also applied for the $e^+e^- 
\ra \bar{t}t+ \cdots$.}. 

\section{Why are the b-quark mass effects are significant in $\z \ra 3 jets$?}
It might seem surprising that at the $\z$-pole, where the relevant
scale is the $\z$-boson mass, effects of the quark masses
can not be neglected. Indeed, if one considers
inclusive width of the decay $\z \ra \bar{b}b$, according to 
the Kinoshita-Lee-Nauenberg theorem there are  no mass singularities, and
the only way $m_b$ enters calculations\footnote{Taking the $\overline{MS}$
running quark mass one includes the principal part of the NLO QCD corrections
to the total width~\cite{dkz90,brs95}.} is in the
ratio $\overline{m}^2_b(M_Z)/M_Z^2$.
Therefore, quark mass effects in the best-measured observable 
for b-quarks are negligibly small $\sim 10^{-3}$.

But the situation is different in more exclusive processes.
Let's consider the decay $\z \ra \bar{b}bg$, which contributes
to the three-jet final state at the LO.
This process has an infrared singularity in the limit 
when the energy of radiated
gluon energy aproaches zero. To make a physical prediction one has
to introduce kinematical cuts. 
In the $e^+e^-$-annihilation this is usually done by applying one of the
so-called jet clustering algorithms~\cite{jets}. The 
phase-space for $\z \ra \bar{b}bg$ is split into two parts, two-jet and three-jet one,
and this separation is defined by the jet-resolution parameter, $y_c$.
Therefore, instead of two scales in the inclusive process, $M_Z$ and $m_b$,
one has here in addition the new scale, $\sqrt{y_c}M_Z$. The jet-resolution parameter
can be rather small, in the range $10^{-2}-10^{-3}$ and effects of the
quark mass appears as $ m_b^2/(\sqrt{y_c}M_Z)^2$
and can reach several percents.

\section{The three-jet observable}
The convenient observable for studies of the mass effects in the three-jet
final state is defined as follows
\bea
\label{eq:r3bd}
& & R_3^{bd}=\frac{\Gamma^b_{3j}(y_c)/\Gamma^b}{\Gamma^d_{3j}(y_c)/\Gamma^d}
\\ \nonumber
& &=1+r_b \left( b_0(y_c,r_b) + \frac{\alpha_s}{\pi} b_1(y_c,r_b) \right)
+{\cal O}(\as^2) 
\eea
where $\Gamma^q_{3j}$ and $\Gamma^q$ are the three-jet and the total decay
widths of the $\z$-boson into a quark pair of flavour $q$,
$r_b=m_b^2/M_Z^2$. Both the LO function, $b_0$, and the NLO function, $b_1$, 
depend on the jet-clustering algorithm.
Note that although the leading $r_b$-dependence is factorized
for convenience, the above expression is not an expansion
in $r_b$. 

This observable has both experimental and theoretical 
advantages. It is a relative quantity, therefore many experimental
uncertainties due to the normalization drop out.
In addition, in the ratios $\Gamma^q_{3j}/\Gamma^q$
the bulk of electroweak
corrections is cancelled.
The double ratio (\ref{eq:r3bd})
was measured by DELPHI~\cite{delphi97}  and used
for the determination of the $m_b$.
In practice one uses normalization with respect to all light flavours, $uds$.
Such quantity then differs from $R_3^{bd}$, mainly due to the contribution
of the triangle diagrams~\cite{hky91}. The difference is, however, very small
numerically. Similar ratio normalized on $udsc$ was considered in~\cite{bbu97}.

Another observable is the differential two-jet rate, $D_2$ defined as
\be
D_2^{bd} =
\frac{[\Gamma_{2j}^b(y_c+\Delta y_c/2)
      -\Gamma_{2j}^b(y_c-\Delta y_c/2)]/\Gamma^b}
     {[\Gamma_{2j}^d(y_c+\Delta y_c/2)
      -\Gamma_{2j}^d(y_c-\Delta y_c/2)]/\Gamma^d}~.
\label{eq:d2bd}
\ee
with $\Delta y_c$ taken to be sufficiently small. The two-jet width is
obtained from the relation:
$\Gamma^q_{2j} = \Gamma^q - \Gamma^q_{3j}- \Gamma^q_{4j}$,
where $\Gamma^q_{4j}$ is the four-jet width.

Different event shapes observables were considered~\cite{no97,bbu97,chrisman}
in the literature as well.

\section{The leading order calculation}
The LO contribution to the decay $\z \ra 3jets$ is given
by the process $\z \ra \bar{b}bg$.
At the LO we can not specify what value of $m_b$
should be taken in the calculations: all 
definitions of the quark mass 
are equivalent (the difference is due to the higher orders in $\as$). 
One can use\footnote{The values of the b-quark masses are taken from 
the recent sum rules and lattice QCD analyses~\cite{lowen} 
of the $\Upsilon$ and $B$ mesons spectra.}, 
for example, the pole mass $M_b \approx 4.6 GeV$
or the $\overline{MS}$-running mass $\overline{m}_b(\mu)$ at any scale 
relevant to the problem, $m_b \le \mu \le M_Z$, with $\overline{m}_b(m_b)
\approx 4.13 GeV$ and $\overline{m}_b(M_Z) \approx 
2.83 GeV$.
\vskip -10mm
\myfigure{0.cm}{0.cm}{6.5cm}{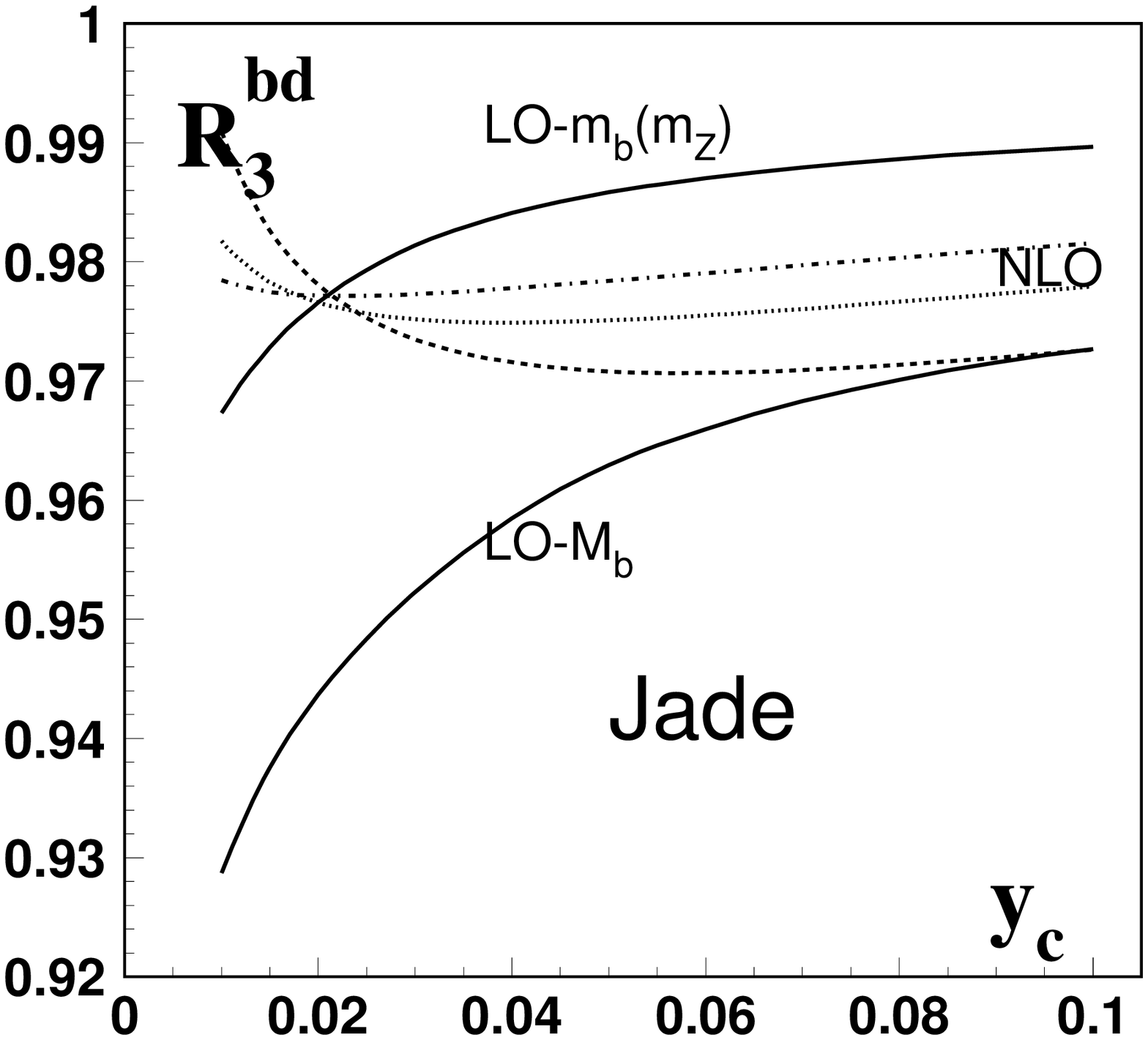}{-0.5cm}
{The ratio $R_3^{bd}$ (Jade) as a function of $y_c$.}{fig:r3fitj}
\vskip -15mm
\myfigure{0.cm}{0.cm}{6.5cm}{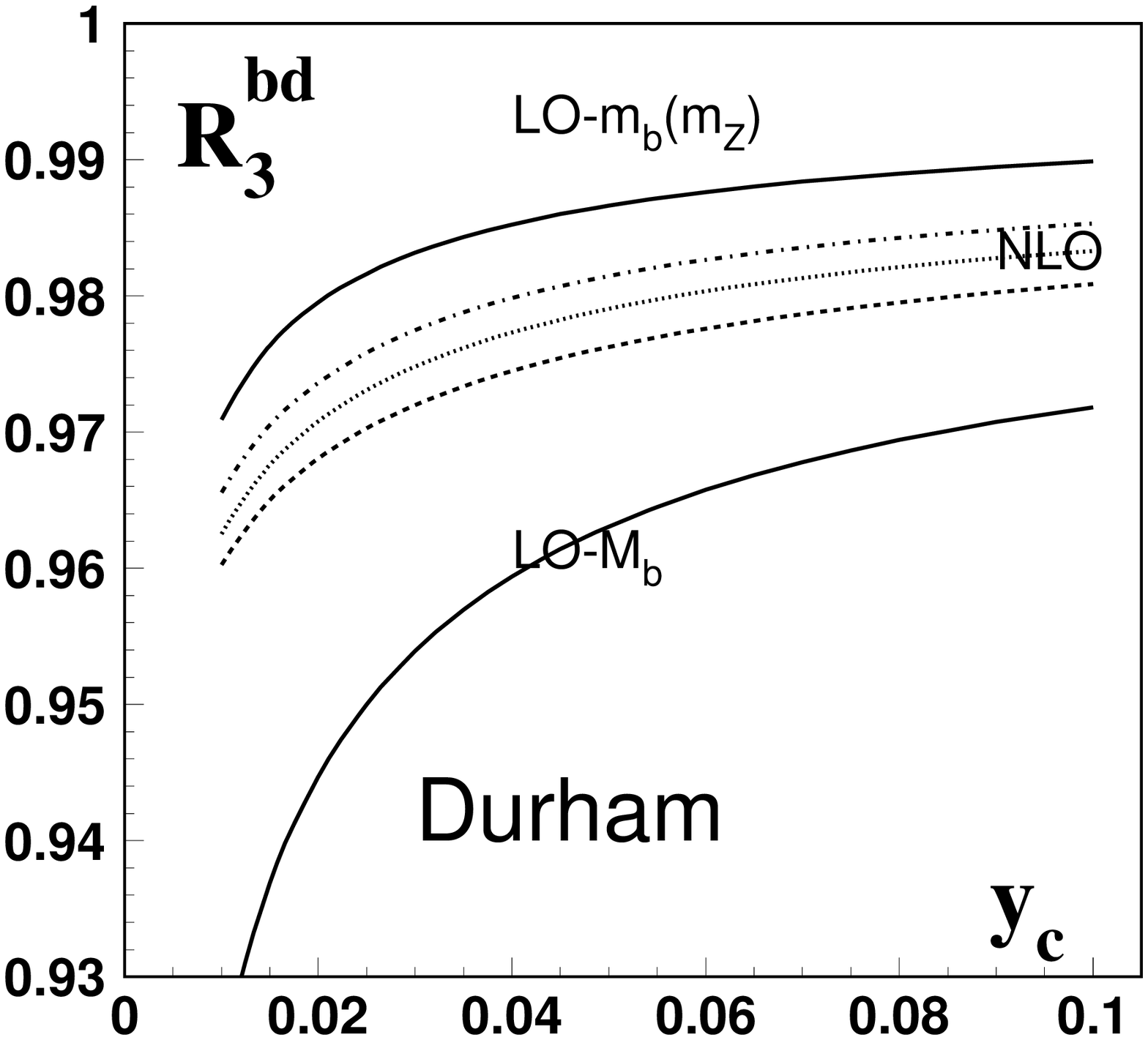}{-0.5cm}
{The ratio $R_3^{bd}$ (Durham) as a function of $y_c$.}{fig:r3fitd}
The results for the ratio $R_3^{bd}$, Eq.(\ref{eq:r3bd}), 
for Jade~\cite{jade} and Durham~\cite{durham} schemes
are presented by solid curves in the Figs.\ref{fig:r3fitj},\ref{fig:r3fitd}.
The upper curve corresponds to $m_b=2.83 GeV$, the lower one to $m_b=4.6 GeV$.
The difference between the two curves gives the size of the uncertainty of  
the LO prediction. To improve the situation
the NLO calculation is necessary.

It is worth mentioning that the residual mass 
dependence in the function $b_0(y_c,r_b)$ is small in the region 
$0.005<y_c<0.1$, especially for the Durham scheme.
The simple interpolation:
$b_0=b_0^{(0)}+b_0^{(1)} \ln y_c+b_0^{(2)} \ln^2 y_c$
can be used in practice~\cite{brs95,brs98}.

\section{The next-to-leading order corrections}
At the NLO there are two different contributions. One comes from one-loop  
corrections to the three-parton decay, $\z \rightarrow b \overline{b} g$.
Another one comes from tree-level four-parton decays $\z \rightarrow
b \overline{b} g g$ and 
$ \z \rightarrow b \overline{b} q \overline{q},~q=u,d,s,c,b$
integrated over the three-jet region of the four-parton phase-space.
The main difficulty of the NLO calculation is the presence of ultraviolet
(UV) and infrared (IR) divergences at the intermediate
stages. The UV divergences in 
the one-loop part are removed by renormalization of the QCD parameters. 
The IR  singularities in the virtual part are due to
massless gluons in the loops. In the four-parton process they appear
when one of the
gluons is soft or two gluons are collinear.
The sum of the virtual and tree-level contributions is, however, IR finite.
In addition,
the finite quark mass makes the NLO calculation technically much
more involved, comparing to the massless one, known since many
years~\cite{ert81,massless}.

The one-loop part is calculated analytically
and dimensional regularization is used to regularize both UV and IR divergences.
The IR singularities appear as simple and double 
poles in $\epsilon,~D=4-2\epsilon$, where
$D$ is space-time dimension. The singular part is proportional to the 
tree-level transition probability and it is cancelled by
the IR-singular part of the four-parton contribution.

There are several methods of analytical cancelation of IR singularities.
In \cite{rsb97,bbu97} the so-called slicing method~\cite{slicing} was used. In this case
the analytical integration over a thin slice at the border of phase-space is performed 
in $D$-dimensions. The integration over the rest of the phase-space, defined
by a particular jet-algorithm, is done
numerically in $D=4$. 
In the third NLO calculation~\cite{no97}, a different approach, the so-called
subtraction method (see e.g. \cite{ert81,subtraction}) was used.

We would like to stress that the structure of the NLO result for the $\z \ra 3jets$
in the massive case is completely different from the massless
one \cite{ert81,massless}. 
In the massive case the collinear divergences 
associated with the gluon radiation from the quarks, are softened
into $\ln r_b$ and only collinear 
divergences due to gluon-gluon splitting remain.
Therefore, the test performed in~\cite{rsb97} 
by recovering the massless limit from the result with finite mass,
is a rather non-trivial one.

In contrast to the LO function, $b_0$, the NLO function
$b_1$ has a significant
residual mass dependence~\cite{rsb97,brs98} 
and the phenomenological 
interpolation, e.g. for Durham scheme, can be chosen as follows
\be
\label{eq:b1}
b_1=b_1^{(0)}+b_1^{(1)} \ln y_c+b_1^{(2)} \ln r_b~.
\ee
The approximation Eq.(\ref{eq:b1}) works well in the region $ 0.01 \le y_c \le 0.1$
and extra powers of $\ln r_b$ and/or mixed terms $ \ln r_b \ln y_c$ do not improve
its quality.

In the NLO calculations one can, and have to, 
specify the quark mass definition.
It turned out that technically it is simpler to use a mixed renormalization
scheme with on-shell definition for the quark mass 
and $\overline{MS}$ definition for the strong coupling. 
In this case observables
are originally expressed in terms of the pole mass. 
This definition of the quark mass can be perfectly used in perturbation theory.
However, in contrast to the pole
mass in QED, the quark pole mass is not a physical parameter. The
non-perturbative corrections to the quark self-energy bring an unavoidable
ambiguity of order $\approx 300 MeV$ (hadron size) to the physical position 
of the pole of the quark propagator~\cite{pole}. 
Above the quark production 
threshold, it is natural to use the running mass definition
(we use $\overline{MS}$). The advantage of this definition is
that $\overline{m}_b(\mu)$ can be used for any scale, $\mu \gg m_b$.
The pole, $M_b$, and the running masses of the quark
are perturbatively related
\be
M_b=\overline{m}_b(\mu)\left[1+\frac{\alpha_s}{\pi}\left(\frac{4}{3}
-\ln\frac{m_b^2}{\mu^2}\right)\right]~.
\label{eq:pole-run}
\ee
Although this relation is known to higher orders in $\as$,
we use its one-loop version to pass from the pole mass to the running one,
which is consistent with  NLO calculations. 

Substituting Eq.(\ref{eq:pole-run}) into Eq.(\ref{eq:r3bd}) we have
\bea
& &R_3^{bd}(y_c,\mb(\mu),\mu)=
\nonumber \\
& &1+\overline{r}_b(\mu)\left[b_0
+ \frac{\alpha_s(\mu)}{\pi} \left(\overline{b}_1
-2b_0\ln\frac{M_Z^2}{\mu^2}\right)\right]
\label{eq:r3bdrun}
\eea
with $\overline{b}_1=b_1+b_0(8/3-2\ln r_b)$ and $\overline{r}_b=\overline{m}_b^2/M_Z^2$.

To relate quark masses at different scales, we 
use the one-loop renormalization group equation, e.g.
$\mb(\mu)=\mb(M_Z)
\left[\frac{\alpha_s(\mu)}{\alpha_s(M_Z)}\right]^{\frac{2\gamma_0}{\beta_0}}$,~
where $\gamma_0 = 2$ and $\beta_0 = 11-\frac{2}{3} n_f$ with $n_f=5$.

\section{Numerical results and discussion.}
In Figs.\ref{fig:r3fitj},\ref{fig:r3fitd} 
we show NLO results, obtained by using Eq.(\ref{eq:r3bdrun}) 
parameterized in terms of $\overline{m}_b(M_Z)=2.83 GeV$ for different values of
the renormalization scale $\mu=10 GeV$ - dashed line, $\mu=30 GeV$ - dotted line
and $\mu=M_Z$ - dashed-dotted line. As expected, the NLO curve for large scale 
is closer to LO curve for $\mb(M_Z)$, while NLO result
 for smaller scale is closer to the LO one for
$M_b$. Like in the massless case~\cite{jets}, 
the NLO corrections for the Jade scheme are larger
than in the Durham and depend significantly on $y_c$ for small values, $y_c <0.025$.

The $\mu$-dependence is a reflection of the fixed order perturbative calculation
and the size of the variation of the observable is usually used to estimate 
higher order corrections.
The study of the $\mu$-dependence for the Durham scheme
is presented in Fig.\ref{fig:r3mud}. 
The $R_3^{bd}$ is shown 
as a function of scale $\mu$ for the fixed value of $y_c=0.02$.
The lower, solid curve is obtained
by using Eq.(\ref{eq:r3bd}), i.e. parameterized in terms of the pole quark mass. In this
case the scale dependence is due to the renormalized coupling constant, $\as$.
Other curves show $\mu$-dependence when
$R_3^{bd}$ is parameterized in terms of the running mass, $\overline{m}_b(M_Z)$,
Eq.(\ref{eq:r3bdrun}), with different definitions of the quark mass 
used in the logarithms: $M_b,~\overline{m}_b(M_Z),~
\overline{m}_b(\mu)$.
\vskip -10mm
\myfigure{0.cm}{0.cm}{6.5cm}{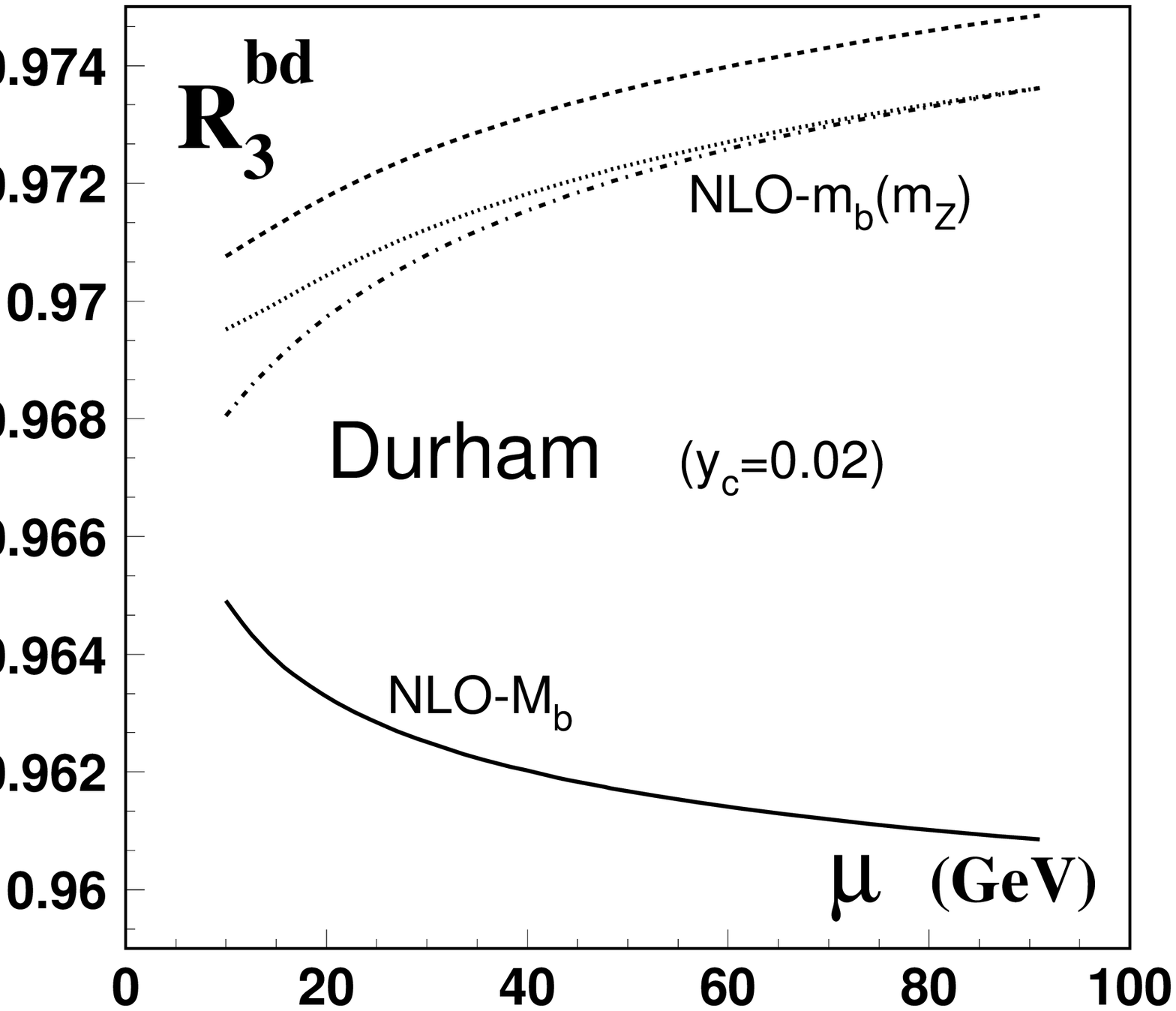}{-0.5cm}
{The ratio $R_3^{bd}$ (Durham) as a function of $\mu$.}{fig:r3mud}
The conservative 
estimate\footnote{The Jade scheme gives significantly larger 
error, see Fig.\ref{fig:r3fitj}.} of the theoretical error for the $R_3^{bd}$
is to take the whole spread given by the curves 
in Fig.\ref{fig:r3mud}.
The uncertainty in $R_3^{bd}$ induces an error in $\overline{m}_b(M_Z)$:
$\Delta R_3^{bd}=0.004 \rightarrow \Delta m_b \simeq 0.23 GeV$,
which is, however, below
current experimental errors, dominated by
fragmentation.

In the Figs.\ref{fig:r3fitc},\ref{fig:d2fitc} 
we show results for the
fixed value of scale, $\mu=M_Z$, for the Cambridge algorithm~\cite{cambridge}, 
a recently proposed 
modification of the Durham scheme. 
\vskip -10mm
\myfigure{0.cm}{0cm}{6.5cm}{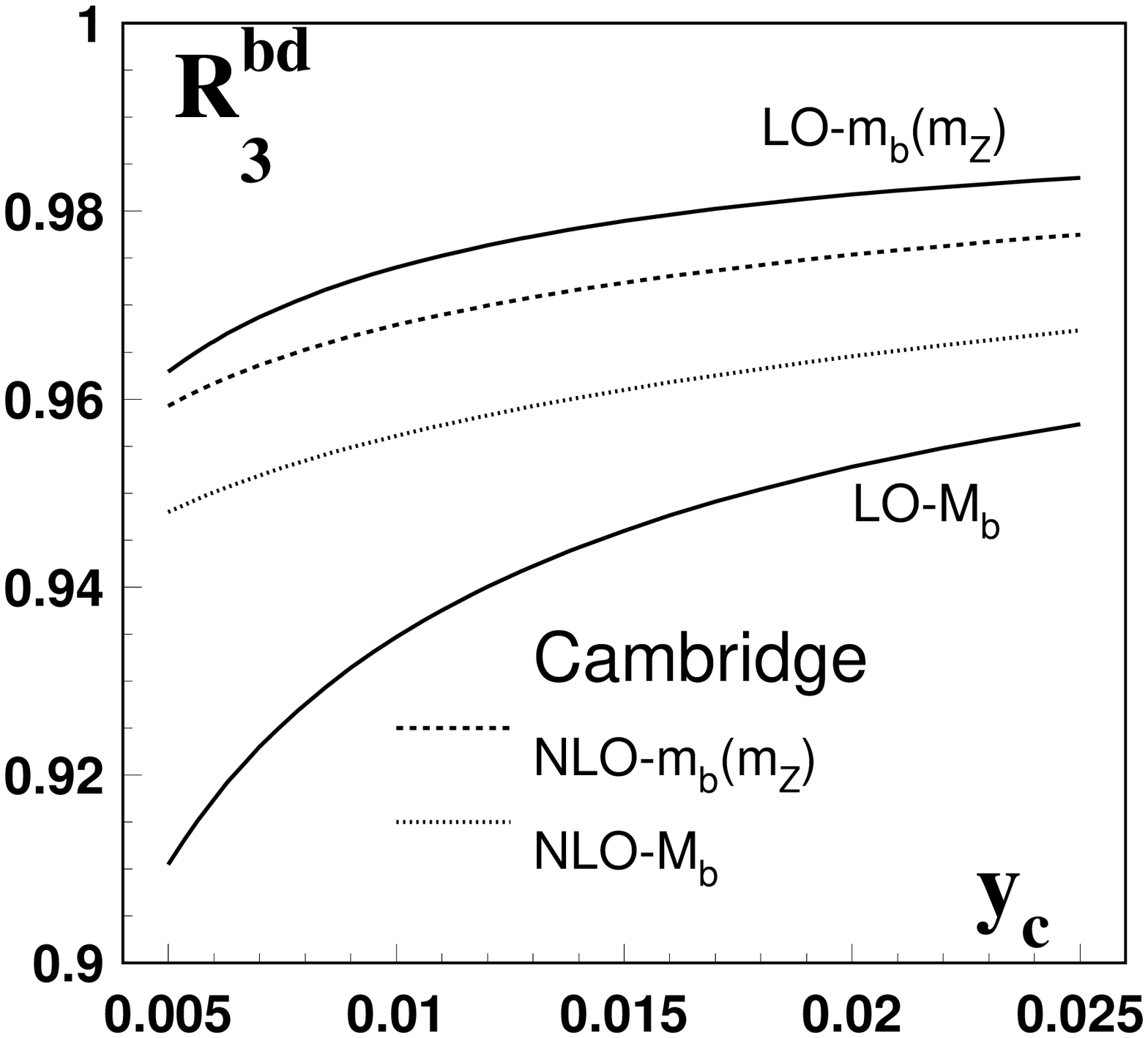}{-0.5cm}
{The ratio $R_3^{bd}$ 
(Cambridge) as a function of $y_c$.}{fig:r3fitc}

It is remarkable that in this scheme and for two observables, $R_3^{bd}$ and 
two-jet differential ratio, $D_2$ (see Eq.(\ref{eq:d2bd})), the LO result
for the running mass, $\overline{m}_b(M_Z)$ is very close to the NLO one.
In a sense, it is similar to 
the total width: the main radiative corrections are taken into
account by the running of the QCD parameters to the $M_Z$-scale.
Note also that for small, but still reasonable value of $y_c \approx 0.01$,
the mass effects in $D_2$ are as large as $10\%$ (although one has to remember
that this is a differential rate and statistical errors are larger here too).
\vskip -10mm
\myfigure{0.cm}{0.cm}{6.5cm}{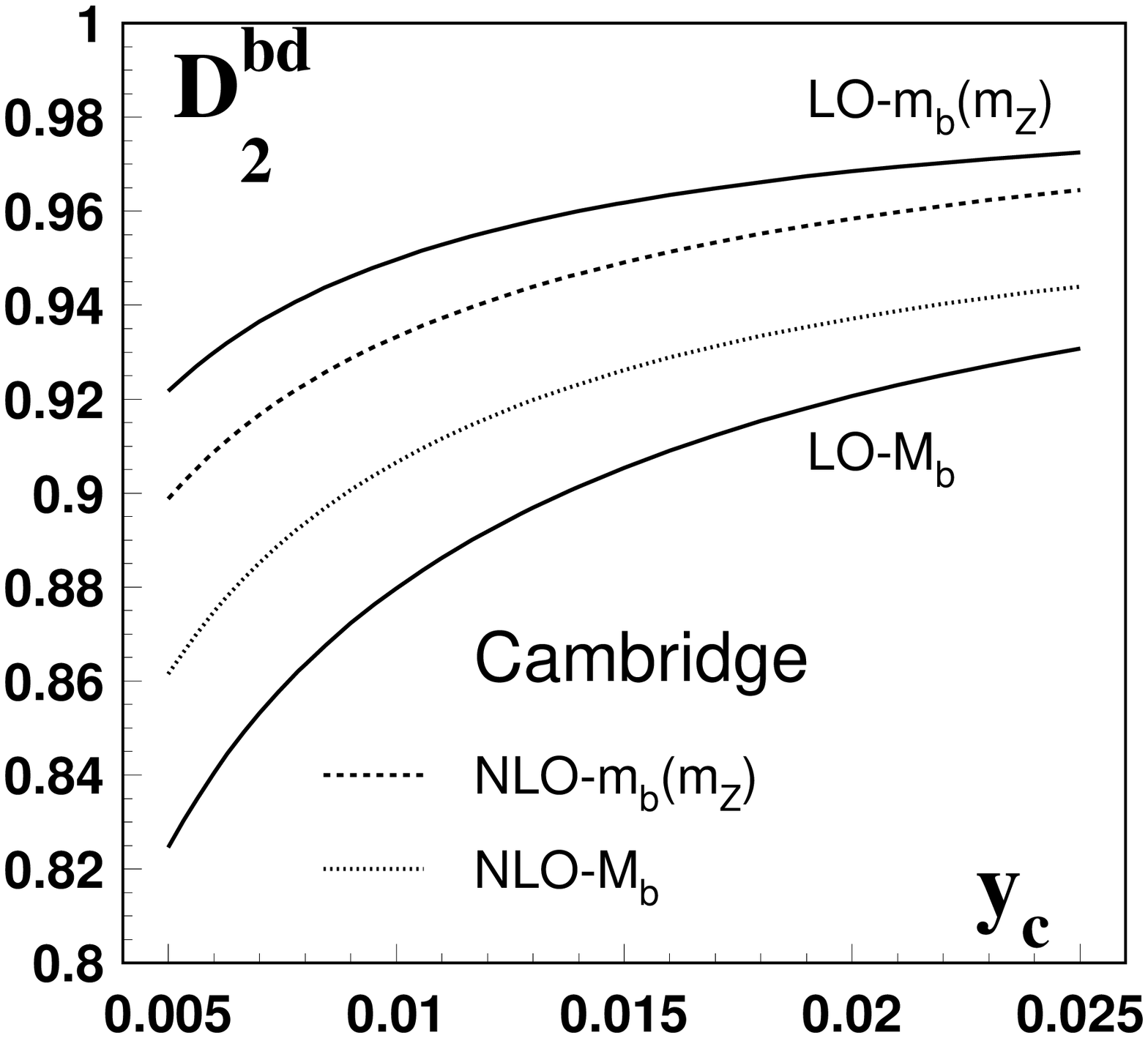}{-0.5cm}
{The
ratio $D_2^{bd}$ (Cambridge) as a function of $y_c$.}{fig:d2fitc}
The scale dependence for the Cambridge scheme
is shown in Fig.\ref{fig:r3muc} for fixed $y_c=0.01$.
The result is also quite remarkable. The ratio expressed in
terms of running mass is very stable with respect to scale variation.
This behavior in the Cambridge scheme is rather promising with respect
to improvements of the DELPHI result~\cite{delphi97} for $\overline{m}_b(M_Z)$.
We refer to~\cite{rsb98} for detailed discussion of the NLO predictions
in this new scheme and to~\cite{juan98} for the first analysis of the
LEP data applying the Cambridge algorithm.
\vskip -10mm
\myfigure{0.cm}{0.cm}{6.5cm}{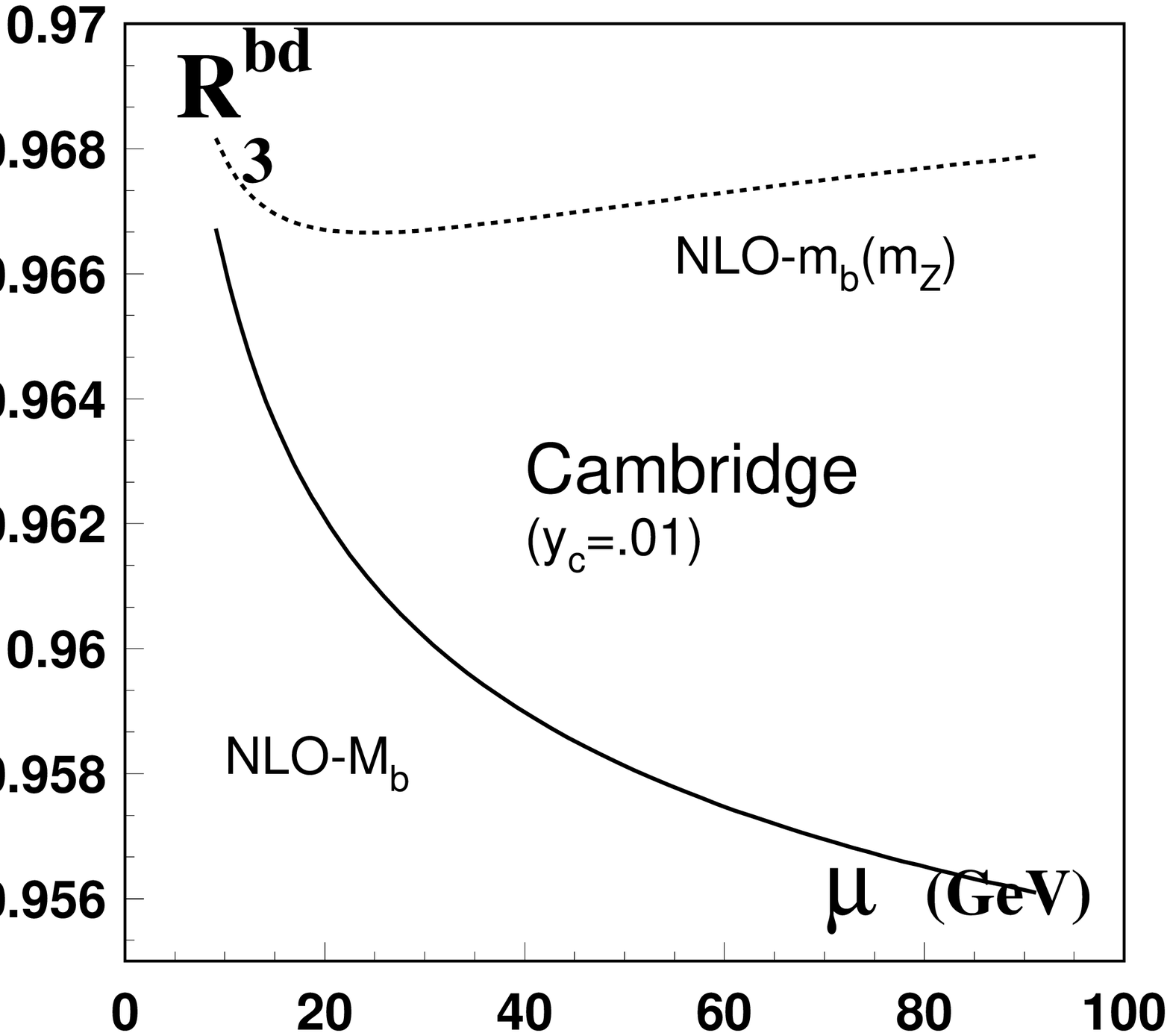}{-0.5cm}
{The ratio $R_3^{bd}$ (Cambridge) as a function of $\mu$.}{fig:r3muc}
\section{Conclusions}
Three-jets observables at the $\z$-peak
(jet rates, differential jet-rates, event-shape observables etc.) 
have significant mass effects ranging up to the $10 \%$ depending on the observable,
jet algorithm
and the value of the jet resolution parameter.
This requires accurate theoretical input, including mass effects, for tests
of the flavour independence
of the strong interactions and measurements
of the bottom-quark mass.

In the last years an important progress was done in the description of the 
decay $\z$ into three-jet with massive quarks. The next-to-leading 
calculations have been done
by three groups and have been successfully used
in the analysis of the LEP and SLC data. 
The NLO corrections are in the range of $1-3\%$
and are within the experimental reach. 
Further studies of different observables and different jet-algorithms 
are oriented on the reduction of the theoretical uncertainty.
One good candidate might be the Cambridge jet-algorithm, 
where the NLO corrections are
particularly small and where the predictions in terms of the running mass, $m_b(M_Z)$
are particularly stable with respect to the variation of the renormalization scale.

\section*{Acknowledgments}
We are very pleased to thank S. Cabrera, 
J. Fuster and S. Mart\'{\i} for an 
enjoyable collaboration.

\end{document}